\documentclass[runningheads]{llncs}
\usepackage{graphicx}
\begin{document}
\title{Extracting Software Requirements from Unstructured Documents}
\titlerunning{Extracting Software Requirements}
%
\author{Vladimir Ivanov\inst{1}
\and
Andrey Sadovykh\inst{1,2}
\and
Alexandr Naumchev\inst{1}
\and
Alessandra Bagnato\inst{2}
\and
Kirill Yakovlev\inst{1}
}
\authorrunning{V. Ivanov et al.}
\institute{Innopolis University, Innopolis, Russia \\ \url{http://university.innopolis.ru} \and
Softeam, France \\ \url{http://www.softeam.com}}
%
\maketitle              
\begin{abstract}
Requirements identification in textual documents or extraction is a tedious and error prone task that many researchers suggest automating. 
We manually annotated the PURE dataset and thus created a new one containing both requirements and non-requirements. Using this dataset, we fine-tuned the BERT model and compare the results with several baselines such as fastText and ELMo. In order to evaluate the model on semantically more complex documents we compare the PURE dataset results with experiments on Request For Information (RFI) documents. The RFIs often include software requirements, but in a less standardized way.
The fine-tuned BERT showed promising results on PURE dataset on the binary sentence classification task. 
Comparing with previous and recent studies dealing with constrained inputs, our approach demonstrates high performance in terms of precision and recall metrics, while being agnostic to the unstructured textual input. 

\end{abstract}
\section{Introduction}
Identifying and understanding technical requirements is a challenging task for many reasons. Requirements come from various sources and in various forms. These different forms may serve different purposes and levels, from initial project proposals down to detailed contracts of individual methods. The forms that appear earlier in the software process are consumed by more business-oriented stakeholders.
Technical people, such as testers, require detailed specifications produced later in the software process to convert them into tests.
The evolving form of describing the initial problem is natural and reflects the process of understanding the problem better with time.
The available requirement datasets reflect those variations. 

One of such is the PURE dataset \cite{8049173} presented by A. Ferrari et al., which includes 79 Software Requirements Specification (SRS) documents. Mainly this set is based on the Web search and includes documents with different peculiarities as well as a lexicon with the widespread writing style of requirements. Authors argue that this dataset fairly fits for various Natural Language Processing (NLP) tasks such as requirements categorisation, ambiguity detection, and equivalent requirements identification, etc. But at the same time, additional expertise, as well as an annotation process, are primarily required. However, some other datasets might be challenging. The Request for Information (RFI) documents in itself specify an initial set of ``best wishes'' for a required solution as well as a variety of constraints in terms of deployment, compatibility or guarantees for maintenance and support. Those documents are meant for engineers though the language used, format, styles and structure are far from being standard. Those parameters largely depend on the organization that issued those documents, personnel involved in designing specifications and traditional practices of the domain the organization belongs to. On the other side, the provider of the solution must quickly qualify the RFI and prepare a well-thought offer that would maximize the value for the customer at minimum costs for the provider. Such analysis is mostly done manually that is costly, time-consuming and error-prone. 

In this paper, we present an approach for extraction of requirements coming from unstructured text sources. We fine-tuned the Bidirectional Encoder Representations from Transformers (BERT) \cite{devlin2018bert} model with PURE-based manually annotated dataset of 7,745 sentences to identify technical requirements in natural text. The ReqExp showed 92\% precision and 80\% recall on the PURE dataset. In order to evaluate BERT, we compared the results with previous state-or-art methods such as fastText and Embeddings from Language Model (ELMo) with an SVM classifier on the same PURE-based dataset, where BERT showed superior results in the precision, but worse results in the recall. Afterwards we carried-out additional experiments with more semantically difficult texts from RFIs. We applied all the candidate models trained on PURE-based dataset to a private dataset of 380 sentences extracted from RFI documents in order to evaluate the possibility to transfer the result to a new domain. BERT again showed superior results in precision (90\%), but inferior in recall (71\%). With a superior recall metrics fastText (82\%) and ELMo (72\%) baseline methods showed less false negatives in detecting requirements, i.e. passed through less of undetected requirements. The results obtained in the limited set of experiments showed optimistic outcomes. We ran an additional experiment with a larger training set, consisted of all the text samples from the PURE and obtained precision of 86\% and recall of 84\% that suggests further fine-tuning of ReqExp on a larger dataset. Moreover, the analysis of related work indicated that the ReqExp approach has certain advantages, since, for example, it is agnostic to the type and the structure of input documents.

Unfortunately, the datasets of comparable studies are not available for a thorough analysis. In order to avoid the same problem with our study, we provide our dataset with manually annotated sentences from the PURE corpus to the community for comparison studies. The initial intention for this study came from to Softeam company from France. They kindly provided the dataset of RFI documents and endorsed our work. Softeam colleagues confirmed that the results are promising for the industry with a high potential to be used in the company's products and processes.

To summarize this paper, the contributions of this work are: (1) adapted PURE-based dataset for classification tasks with manually annotated sentences containing both requirements and non-requirements; (2) the description of BERT-based tool chain applying NLP for requirements extraction; (3) the results of evaluation on SRS sentences from the PURE dataset in comparison with baseline methods -- fastText and ELMo; (4) the results of evaluation on a provide dataset of RFI documents; (5) analysis of comparable studies. Therefore, the current paper starts with the outline of the ReqExp approach in Section \ref{approach} describing the conceptual method and the dataset construction for profiling BERT neural network, followed by classifier models details in Section \ref{baselines} , continued by validation of the prototype on PURE-based dataset and real RFI documents in Section \ref{validation}, to be concluded by a comparison with the related work in Section \ref{sota}.

\section{Methods and Datasets} \label{approach}

\begin{figure}
    \centering
\includegraphics[width=6cm]{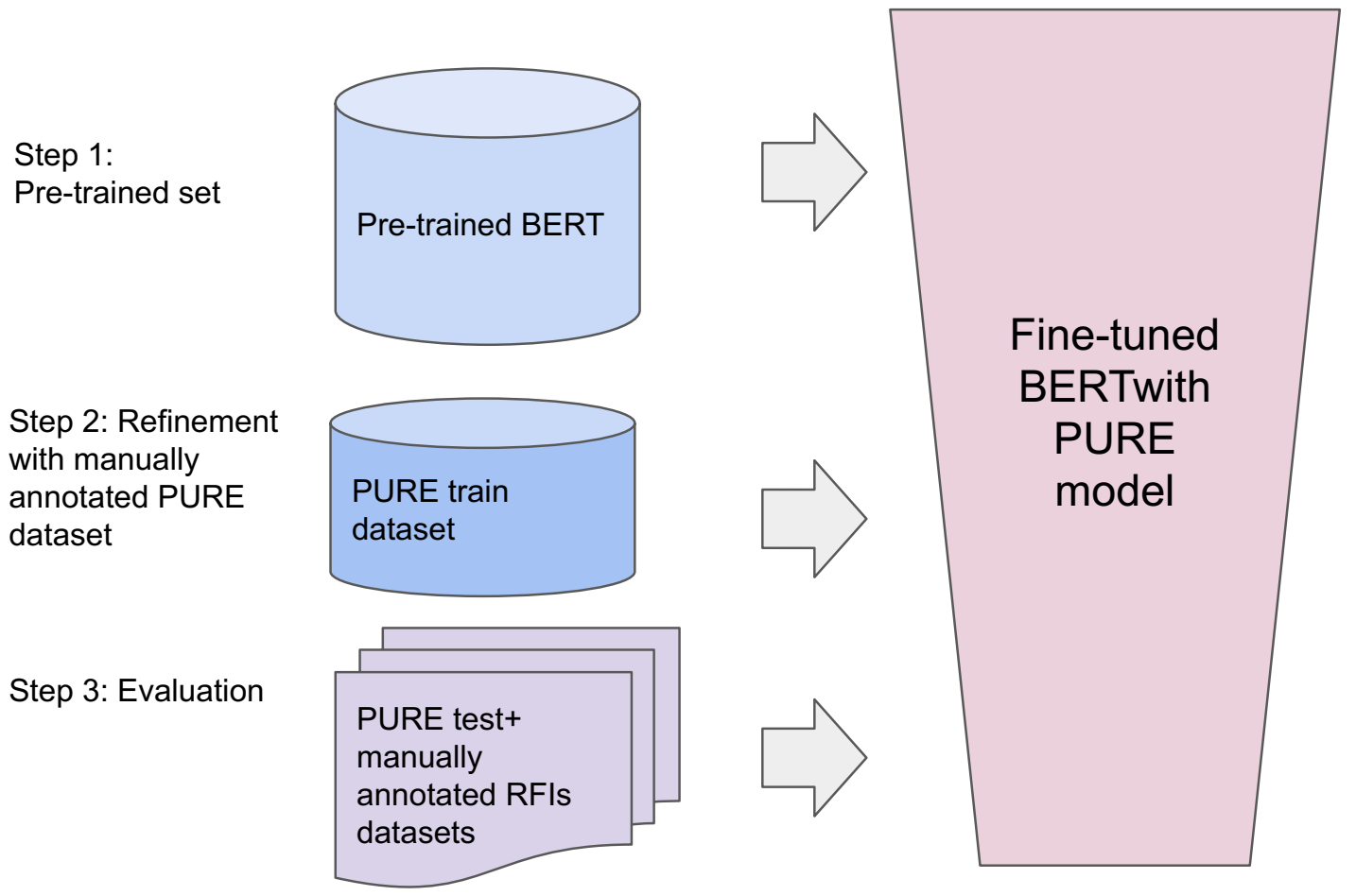}
    \caption{The training process to obtain a model for binary sentence classification for requirements extraction.}
    \label{fig:proto_training}
\end{figure}

In this section we describe the process of building a binary sentence classifier. 
This process has several steps presented in Figure \ref{fig:proto_training}.
The ultimate goal was to efficiently build a training dataset that can be used to fine-tune the BERT model and compare it with other advanced NLP methods. Bidirectional Encoder Representations from Transformers (BERT) is a  model by Google pre-trained on a huge text corpus\cite{devlin2018bert}. BERT was tested on multiple NLP tasks and achieved state-of-the-art performance. A common way to use BERT for a domain-specific task, such as software requirements extraction, is to fine-tune the model i.e., to optimize model's weights on a domain-specific dataset. To solve a binary classification problem, such dataset should contain two subsets of sentences: (i) a subset of sentences that contain requirements, and (ii) a subset of sentences that does not contain requirements. However, there were no such dataset readily available for requirements extraction.

Thus, the first step of our approach is creating the domain-specific training dataset for fine-tuning. To speed up the
process of corpus creation, we decided to process an existing corpus of software specification documents. 
In \cite{8049173}, Ferrari et al. presented a text corpus of public requirements documents (PURE)\footnote{Data of the corpus can be found here: https://zenodo.org/record/1414117}. The dataset contains 79 publicly available natural language requirements documents. We applied PURE corpus as our main source of both requirements and non-requirements for designing our own dataset. Initially, documents were presented in the form of raw text with different format, structure and writing style. 

At the beginning we made an attempt to extract sentences with requirements using some parsing and extraction engines. Unfortunately, this idea was rejected due to unsatisfactory quality of extracted text samples. Despite that authors did provide a set of text requirements in the form of \textit{.json} files to simplify the extraction process, this was not enough to design a representative subset for each class. It was decided to process documents manually preserving initial syntax and semantics in the extracted text samples. Still this process was associated with some challenges, which required us to apply the following conventions:
\begin{itemize}
    \item Due to different length and structure of requirements, it was decided to assume one sentence as one data unit. It means that a requirement, which consists of several sentences, was divided into several units.
    \item Sometimes requirements were written in the form of a list with many elements. In such case, this list was transformed into one sentence separating parts by commas. In some rare cases, where elements were individual sentences, such requirement was divided into several sentences accordingly.
\end{itemize}

Overall, we extracted 7,745 sentences, where 4,145 were requirements and 3,600 were non-requirements from 30 documents. The extraction was done from the appropriate sections of the documents. Usually each document provides structural elements like table of elements, requirements-focused sections with appropriate names or contextual footnotes. The latter one might have some specific numbering used in a document and initially explained by the authors, or lexical elements that are usually inherent for requirements such as modal verbs like 'must', 'shall', 'should', etc. When identifying requirements and non-requirements we followed principle described in \cite{Abualhaija2019}. 

Sentences with requirements were manually extracted by one specialist in Software Engineering area and then independently validated by another expert from our research group. 
Some special cases were separately discussed in the group to make annotation more consistent. In the case when there was no agreement concerning which label should be utilized to those text samples, they were removed to maintain consistency of the set.

\section{Models for Sentence Classification}
\label{baselines}
For this study we chose two baseline approaches, which preceded the BERT architecture, for binary sentence classification. Specifically we considered fastText model and ELMo embeddings with a classical machine learning classifier. The latter one was chosen through an experiment stage to choose the most promising model based on the most explicit and informative metrics during comparison.

\subsection{fastText baseline model}

fastText is a method for constructing word embeddings based on the Word2Vec principle, but with some extension. An embedding for each word is learnt from embedding vectors for character n-grams that constitute the word. Thus, each word vector is represented as the sum of the n-gram vectors. This extends the Word2Vec type models with subword information. 
Once a word is represented using character n-grams, a skipgram model is trained to learn the embeddings. This approach provides a fast and scalable solution, allowing to train models on large corpora \cite{fasttext}.
For the down-stream task (binary classification of sentences) we use a linear classifier that is default (a multinomial logistic regression) in the fastText library.

An important part in designing predictive models is associated with finding the best hyper-parameters. Usually, it is a cumbersome and time-consuming task, especially when doing it manually. For this purpose, we applied an autotune feature located inside the fastText functionality by using the validation set presented earlier \footnote{https://fasttext.cc/docs/en/autotune.html}.

\subsection{ELMo with SVM classifier}

Embeddings from Language Models, or ELMo is an advanced method for constructing a contextualized word representation. The core ability is to model different characteristics of word use together with linguistic context. Specifically, these representations are internal states of a deep bidirectional language model that is pre-trained on a large corpus. ELMo maximizes the log-likelihood of the forward and backward directions in a sequence of words \cite{ELMo}. These produced vectors can be applied as features for various NLP purposes.

To make a decision for distinguishing among sentences with requirement and without, one needs to use a classifier on top of the features extracted from text. To find the best classifier, we compared several popular machine learning models such as Logistic Regression, Random Forest classifier, Support Vector Machines and Multi-Layer Perceptron. To this end the Grid Search was applied with specific set of initially predefined hyperparameters for each model. The process of assessment was based entirely on the training set. As a result we chose with Support Vector Machines with polynomial kernel that showed the highest result in terms of F1-score. The results related to this baseline will be marked as ``ELMo+SVM'' in the following sections.

\subsection{BERT-based approach}
Recently using a large pre-trained language model has shown a breakthrough in many tasks of natural language processing including text classification. The baseline approach to transfer learning typically implies pre-training a neural network on a large corpus of texts with further fine-tuning the model on a specific task. In this study, we use pre-trained BERT \cite{devlin2018bert} model\footnote{Specifically, we use the BERT-base uncased model from \url{https://github.com/google-research/bert}} as a baseline. In all experiments with BERT, we adopted the implementation of fine-tuning process for binary text classification presented in \cite{sun2020finetune}. Hyper-parameters used in fine-tuning are the following: batch size is 64, learning rate is $2\cdot 10^{-5}$, max sequence length is 100, number of epochs is 2.
Overall schema is presented in Fig.  \ref{fig:proto_training}.

\section{Experimental Setup and Validation} \label{validation}

\subsection{Experimental setup}
Empirical evaluation of model performance was carried out in two phases. First, we used a classical approach with splitting dataset into  disjoint sets of sentences (namely, `train', `test' and `validation' sets). All models were trained on the same `train' subset and tested on the same `test' part from the PURE dataset. We did not apply cross-validation due to the special rule of splitting collection into train and test sets, i.e. \textit{sentences from some document should appear in either the train set or in the test set}. The dataset as well as the split is available for the sake of reproducibility and comparison of results.

Second, we used an out-of-sample data for evaluation of the classifiers. To this end we tested the performance on sentences manually extracted from the RFI documents. All models were trained on the same `train' part and tested on the same group of sentences extracted from RFI. We argue that such experimental conditions are more relevant in practice. Indeed, we expected decreasing of the performance on the RFI documents. This shows that fine-tuning of large pre-trained language models is not a panacea in practice, however it gives a viable alternative to rule-based tools. Finally, we show that adding more data on the fine-tuning phase leads to better performance on the out-of-sample data.

\subsection{Train, test and validation sets}

As for the data split, we had 7,745 sentences extracted from the PURE dataset to split them up into train, test and validation subsets by the parts of approximately 70\%, 20 \% and 10\% accordingly. In order to consistently separate our dataset, sentences from every document were fully added only to one specific subset. We found a particular combination of documents for each subset that preserve the stated separation and at the same time preserves the balance of class labels. Statistics of the split is in sentences is the following: train set (5306); test set (1534);  validation set (905).
The dataset is available at \url{https://zenodo.org/record/4630687}.

\subsection{Experiments with PURE documents}
The main purpose of this study was about assessing three major classification metrics: Precision, Recall, and F1-score. Those metrics are perceived as the golden standard in Machine Learning and Deep Learning areas.

Result of the first phase of evaluation is presented in Table \ref{tab:tbl-exp}. As it was expected, more advanced model (BERT) showed better results in terms of F1-score. BERT-based model showed high precision (0.92) and lower recall (0.8). Results of the BERT model is available at \url{https://bit.ly/3oPElMm}. However, the values of precision and recall metrics behave differently for the fastText and ELMo-based baselines. fastText-based classifier showed better Recall (0.93) comparing with other architectures. This property might be useful in some cases when it is necessary to extract more relevant sentences and text patterns associated with requirements. 

\begin{table}
    \centering
    \caption{Results of experiments with PURE}
    \label{tab:tbl-exp}
    \begin{tabular}{l|c|c|c|c|c|c|c}
\textbf{Model} & \textbf{F1} & \textbf{P} & \textbf{R} & \textbf{TP} & \textbf{TN} & \textbf{FP} & \textbf{FN}\\
    \hline fastText & .81 & .72 & \textbf{.93} & 763 & 419 & 295 & 57\\
    \hline ELMo+SVM & .83 & .78 & .88 & 827 & 364 & 231 & 112\\
    \hline 
    BERT & \textbf{.86} & .92 & .80 & 841 &	407 &	69 &	217 \\
    \end{tabular}
\end{table}

\subsection{Experiments with RFI documents}
In order to further evaluate our approach on less standardized requirements documents, we applied our prototypes to five anonymized documents provided by Softeam, France, as issued by their customers. 
We parsed the obtained documents to extract the paragraphs and then triggered the automatic requirements extraction from those paragraphs. After that we manually annotated all the sentences and thus assessed the four major metrics - the number of the true positives, the true negatives, the false positives and the false negatives. Then the other metrics such as precision, recall and F1-score were derived. The results can be found in Table \ref{tab:tbl-exp2}. 

To the validity of the results, it should be noted that the annotation was done by one single specialist who was aware of the context of the Request for Information (RFI) documents. This choice is justified to ensure the uniformity of the manual annotation, since we noticed that the understanding of whether a sentence is a requirement depends on context very much. In addition, it should be mentioned that our attempts of manual annotation by junior researchers were unfruitful since the results were completely unsatisfactory.

\begin{table}
    \centering
    \caption{Results of experiments with RFI}
    \label{tab:tbl-exp2}
    \begin{tabular}{l|c|c|c|c|c|c|c}
\textbf{Model} & \textbf{F1} & \textbf{P} & \textbf{R} & \textbf{TP} & \textbf{TN} & \textbf{FP} & \textbf{FN}\\
    \hline fastText & .65 & .54 & \textbf{.82} & 146 & 83 & 120 & 31 \\
    \hline ELMo+SVM & .69 & .66 & .73 & 177 & 48 & 89 & 66 \\ \hline
   
   BERT & \textbf{.80} & .90 &	.71 & 190 &	93 &	21 &	76	 \\
   
    \end{tabular}
\end{table}



We also faced several challenges when were manually annotating the sentences from RFIs. First, RFI documents often specify technical requirements in the form of a question: ``Will your solution provide a particular feature?'' These questions are difficult to qualify as technical requirements in traditional sense. 
Second, a large part of RFI documents concern the constraints for the submission of the response such as delay, length, and language of the response. Should that be qualified as a requirement to documentation? All these issues are postponed to the follow up studies where we intend to refine the results. In particular, the study triggered more reflection about the nature of the requirements, since the specification practice is far from the form specified in the traditional requirements engineering standards.

Finally, we ran an additional experiment when we fine-tuned the best model on the sentences from train, validation and test sets altogether with further evaluation on the sentences from RFI documents. All parameters of fine-tuning were the same as in the first set of experiments. The performance has grown significantly (P=0.86, R=0.84, F1=0.85) compared to the initial setup with the fine-tuning on the train set only (Table \ref{tab:tbl-exp2}). Thus, this indicates a possibility that the quality of the result can be improved further by improving and augmenting train dataset.

\section{Related Work} \label{sota}
The related work analysis in the present section relies on a recent mapping study of natural language processing (NLP) techniques in requirements engineering \cite{zhao2020natural}. 
The research group analyzed 404 primary studies relevant to the NLP for requirements engineering (NLP4RE) domain.
They mapped these studies into several dimensions and identified the key categories along each dimension. Here, we focus on the \emph{Research Facet} that enumerates categories of research and of evaluation methods.
Our contribution falls into the \emph{solution proposal} category: we develop a new solution (a tool) for identifying likely requirements in software-related texts.
We used the following methods for evaluating our proposed solution:
\begin{itemize}
    \item \emph{Experience Report}: ``the result has been used on real examples, but not in the form of case studies or controlled experiments, the evidence of its use is collected informally or formally.''
    \item \emph{Laboratory Experiment with Software Subjects (LESS)}: ``a laboratory experiment to compare the performance of newly proposed system with other existing systems.''
\end{itemize}






Below we list studies from \cite{zhao2020natural} that fall into the same categories as the ReqExp approach does.
Some of the studies make strong assumptions about the respective input and/or output artifacts.
We do not compare ReqExp with these studies because we do not make such assumptions: ReqExp takes as input arbitrary technical documents and identifies in them statements that look like potential system requirements.
The approaches that work with documents from later stages of the requirements process are:
\begin{itemize}
    \item ARSENAL \cite{ARSENAL} deals with detailed behavior specifications taking the form of ``shall'' statements and automatically converts them into formal models.
    \item AutoAnnotator \cite{AutoAnnotator} assumes to receive requirement documents as input and assigns semantic tags to these documents.
    \item CHOReOS \cite{CHOReOS} works with well-formed requirements specified by domain experts.
    \item Doc2Spec \cite{Doc2Spec} works with JavaDoc and other in-code documentation.
    \item ELICA \cite{ELICA} extracts and classifies requirements-relevant information based on domain repositories and either existing requirement documents or elicitation transcripts.
    \item FENL \cite{FENL} extracts software features from online software reviews.    
    \item GUEST \cite{GUEST} works with well-formed requirement documents to extract goal and use case models.
    \item GaiusT \cite{GaiusT} applies to regulatory documents and extracts statements of specific types with predefined structures.
    \item NLP-KAOS \cite{NLP-KAOS} analyzes research abstracts and produces KAOS goal diagrams from them.
    \item SAFE \cite{SAFE} analyzes descriptions and user reviews of mobile apps in application stores and extracts features from them.
    \item SNACC \cite{SNACC} automatically checks design information for compliance with regulatory documents.
    \item Text2Policy \cite{Text2Policy} processes requirements in natural language to search statements of a predefined type (access control policies).
    \item UCTD \cite{UCTD} takes use cases as input and produces use cases enriched with transaction trees.
    \item Guidance Tool \cite{GuidanceTool} works with texts that already look much like use cases and transforms them into properly structured use cases.
    \item Text2UseCase \cite{Test2UseCase} works with texts that already describe scenarios and transforms them into well-formed use cases.
\end{itemize}


The recent study \cite{Abualhaija2019} has explored the very related topic of demarcating requirements in textual specifications with ML-based approaches. The authors applied similar methods on a private dataset of 30 specification from industrial companies with presumably 2,145 sentences containing 663 requirements and 1,482 non-requirements candidates. They obtained comparable average precision of 81.2\% and an average recall of 95.7\%. Unfortunately, it impossible to compare our results on the same dataset. However, we should underline the major differences such as (1) our publicly available dataset is much larger and more balanced; (2) we trained and validated the model on completely different documents; (3) we applied much simpler end-to-end training method without any costly feature engineering; (4) that way our method is arguably more suitable for documents with complex semantics such as RFIs.

Another similar study \cite{norbert} has presented an application of BERT for multiclass classification of requirements. Authors were focused on the following subtasks:
\begin{itemize}
    \item Binary classification of functional (F) and non-functional (NFR) requirements on the original  dataset.
    \item Binary and multiclass classification of the four most frequent NFR subclasses on all NFR in the original dataset.
    \item Multiclass classification of all NFR subclasses on NFR in the original NFR dataset.
    \item Binary classification of requirements based on functional and quality aspects using the relabeled NFR dataset by another study.
\end{itemize}

For each subtask they provided classification results in terms of precision, recall and F1-score. As it was expected, BERT showed the most promising results comparing with other baseline approaches. 

\section{Conclusions} \label{conclusions}

Addressing the customers' needs involves a huge number of documents expressing requirements in various forms. In the initial steps of those interactions with customer the requirements are provided in RFI documents.
Those documents often combine technical and commercial requirements as well as provide other constraints. 
The formats and styles of RFI vary a lot, which makes it difficult to technically qualify the needs and timely prepare an offer.  

In this paper we propose a requirements extraction approach based on Machine Learning. We used BERT as the basis of the ML model, which we fine-tuned with a manually annotated dataset constructed from an existing requirements specification repository. We have compared the model to two competitive baselines (fastText-based and ELMo-based). Indeed, the study is preliminary as we just compare baselines models and do not propose any specific improvements for the architectures. However, the results of the baselines are quite promising and can be used in industry.

We validated our approach on five documents coming from Softeam company. The approach and validation results were compared to our previous studies. The five documents were annotated automatically at the sentence level (in total, 380 sentences with 6,676 tokens were extracted) and then evaluated by an expert to calculate the performance scores. This analysis showed that our approach superior, since it supports unconstrained textual specifications - it is agnostic to the format of the input textual document. The performance characteristics of our approach are on the comparable level even without costly feature engineering though a progress can be made. We may further evolve the prototype if, for example, we continue the fine-tuning. In addition, by combining train, test and validation sets into one training set has shown a dramatic increase of BERT performance on more semantically complex RFI documents (P=0.86,  R=0.84, F1=0.85). This suggests usefulness of further work on manual annotation of the remaining documents from the PURE dataset and augmenting the train set.

The collaboration with Softeam made us realize the industrial usefulness of the approach in real cases as it can improve current tools and service by that company. Overall, the obtained results motivate us to continue further evolve our approach. In fact, the improvement will include (i) using a corpus of software-related texts for further pre-training, (ii) fine-tuning models on a multi-task datasets and (iii) annotating more data with requirements.

\section*{Acknowledgments} 

This research has been financially supported by The Analytical Center for the Government of the Russian Federation (Agreement No. 70-2021-00143 01.11.2021).

\bibliographystyle{IEEEtran}
\bibliography{sample-bibliography}





\end{document}